\documentclass[twocolumn,letter,showpacs,preprintnumbers,amsmath,amssymb]{revtex4}

\usepackage{epsfig}
\usepackage{graphicx}
\usepackage{dcolumn}
\usepackage{bm}

\def\be{\begin{equation}}
\def\ee{\end{equation}}
\def\bea{\begin{eqnarray}}
\def\eea{\end{eqnarray}}
\def\ie{{\it i.e. }}

\def\k{\kappa}
\def\cl{C_{\ell}}

\def\nl{N_\ell}

\def\nhat{\hat{\bm{n}}}
\def\fsky{f_{\mathrm{sky}}}

\begin{document}

\preprint{APS/123-QED}

\title{Next Generation Redshift Surveys and the Origin of Cosmic Acceleration}

\author{Viviana Acquaviva}
\email{vacquavi@princeton.edu}
 \author{Amir Hajian}
\email{ahajian@princeton.edu}
\author{David N. Spergel}
\email{dns@astro.princeton.edu}
\author{Sudeep Das}
\email{sudeep@astro.princeton.edu}
\affiliation{Department of Astrophysical Sciences, Peyton Hall, Princeton University, Princeton, NJ 08544}

\date{\today}

\begin{abstract}
 Cosmologists are exploring two possible sets of explanations for the remarkable observation of cosmic acceleration: dark energy fills space or
general relativity fails on cosmological scales. We define a null test parameter $\epsilon(k,a) \equiv\Omega_m^{- \gamma} d \ln D / d \ln a - 1$, where $a$ is the scale factor, $D$ is the growth rate of
structure, $\Omega_m(a)$ is the matter density parameter, and $\gamma$ is a simple function of redshift.
 We show that it can be expressed entirely in terms of the bias factor, $b(a)$, (measured from cross-correlations with CMB lensing) and 
the amplitude of redshift space distortions, $\beta(k,a)$.  Measurements of the CMB power spectrum determine $\Omega_{m\,0} H_0^2$.
If dark energy within GR is the solution to the cosmic acceleration problem,
then the logarithmic growth rate of structure $d \ln D / d \ln a = \Omega_m^\gamma$. Thus, $\epsilon(k,a) =0$ on
linear scales to better than 1\%.  We show that  in the class of Modified Gravity models known as $f(R)$, the growth
rate has a different dependence on scale and redshift.  
 By combining measurements of the amplitude of $\beta$ and of the bias, $b$,  redshift surveys will be able to determine
the logarithmic growth rate as a function of scale and redshift.  We estimate the predicted sensitivity
of the proposed SDSS III (BOSS) survey and the proposed ADEPT mission and find that they will test structure growth in General Relativity to the percent level.
\end{abstract}

\pacs{}
\maketitle

\section{Introduction}
\label{intro}
 In General Relativity (GR), there are four variables characterizing linear cosmic perturbations: the two gravitational potentials $\Psi$ and $\Phi$, the anisotropic stress, $\sigma$, and the pressure perturbation, $\delta p$. All of these variables can depend on the wave number, $k$, and the expansion factor $a$. 
We focus initially on models with no dark energy clustering or pressure and discuss these effects later in the paper.
We assume scalar linear perturbations around a flat FRW background in the Newtonian gauge,
\be
ds^2 = -(1+2\Psi) dt^2 + a^2(1+2\Phi)dx^2
\ee
 and work in the quasi-static, linear approximation, which is valid for sub-horizon modes still in the linear regime. \\
 The evolution of perturbations is described by the continuity, Euler and Poisson equations (\ie \cite{Jain:2007yk}):

\bea
\Delta'_m &=& -k_H V_m, \\
V'_m + V_m &=& \frac{k}{aH} \Psi, \\
k^2 \Phi  &=& - 4\pi G a^2 \rho_m \Delta_m.
\eea
With the assumption of no anisotropic stress, $\Phi = - \Psi$, these equations can be combined to derive the equation of motion for the growth factor $D$, defined as $ D(k,z) = \delta_m(k,z)/\delta_m(k,z=\infty)$:
 \be
\label{growth}
D'' + (2 + \frac{H'}{H})D' - \frac{4\pi G}{H^2}\rho_m D = 0;
\ee
where $H = \dot{a}/a$ is the Hubble function, and a prime denotes derivative with respect to $\ln a$. \\ 
From Eq. (\ref{growth}) one can infer the two key features of GR with smooth dark energy. First, the growth factor is exactly determined once the Hubble function $H(a)$ is known; and second, since none of the coefficients is a function of scale, the growth factor is scale-independent. Therefore, for a given expansion history, one can test GR in two ways: checking that the theoretical solution of Eq. (\ref{growth}) agrees with observations, and testing the hypothesis of scale-independence \cite{Bertschinger:2008zb,Bertschinger:2006aw,Ishak:2005zs,Daniel:2008et,Dore:2007jh,Linder:2007hg,Zhang:2007nk,Wang:2007ht,Sahni:2006pa,Yamamoto:2008gr} . 
 The growth rate of structure in GR is well approximated by  $\Omega_m^\gamma$, where the fitting function $\gamma(z) \simeq 0.557 - 0.02 z$ is accurate at the $0.3\%$ level  \cite{Polarski:2007rr}.
 We define a function that tests the growth rate of structure and can be directly related
 to observables:
\be
\epsilon(k,a) = \Omega_m^{-\gamma(a)} \frac{d\ln D}{d\ln a} -1  = \frac{a^{3\gamma} H(a)^{2 \gamma} }{(\Omega_{m,0} H_0^2)^{\gamma}}\frac{d\ln D}{d\ln a} -1 .
\label{eq:eps}
\ee
The combination $\Omega_{m,0} H^2_0$ can be constrained via CMB measurements; it is currently known to within the 5\% level fro the WMAP5 data \cite{Dunkley:2008ie}, with an expected gain of a factor $\simeq 4$ from the upcoming satellite CMB mission Planck \cite{Planck}. The solid line in Fig. \ref{fig:growth} shows $\epsilon(a)$ in GR with dark energy: regardless of the details of the dark energy model, $\epsilon(a) \simeq 0$ in the linear regime. By measuring this quantity, we can characterize deviations from General Relativity.  \\

\section{f(R) theories in the PPF formalism}
\label{PPF}
\begin{figure}
\includegraphics[width=\linewidth]{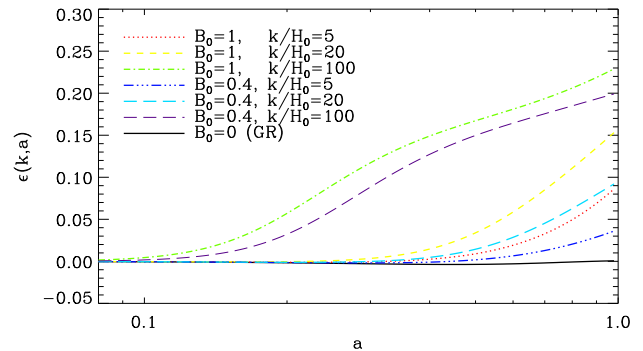}
\caption{The behavior of $\epsilon(k,a)$ = $\Omega_m^{- \gamma} d\ln D/d\ln a - 1$ in GR  (solid line) and in $f(R)$ models, as a function of $B_0$ and $ k$.  Growth is enhanced for $B_0 \ne 0$ and at smaller scales in alternative theories.  In GR, $\epsilon (a) = 0$.}
\label{fig:growth}
\end{figure}

To quantify the expected deviations, we study a class of modified gravity (MoG) models known as $f(R)$ theories, whose action is written as
\be
\mathcal{S} = \frac{1}{16 \pi G}\int d^4 x \sqrt{-g} (R + f(R)) + \mathcal{S}_m;
\ee
$S_{m}$ is the action of standard matter fields. It has been noted long ago \cite{Carroll:2003wy,Capozziello:2003tk} that models where $f(R)$ is an inverse power of the Ricci scalar can give rise to late-time acceleration; however, many of those models have been shown not to be cosmologically viable due to gravitational instability \cite{Dolgov:2003px}. \\
Recently, Song, Hu and Sawicki \cite{Song:2006ej} introduced an effective parametrization for $f(R)$ theories which does not rely on any particular model and is able to discard models suffering from instabilities. The cosmological evolution is obtained by fixing the expansion history to match that of a dark energy model, for which we assume a constant equation of state $w$:
\be
H^2 = H^2_0 (\Omega_m a^{-3} + (1-\Omega_m) a^{-3(1+w)}).
\ee
Such requirement for $H(a)$ translates into a second-order equation for $f(R)$, which can be solved numerically. Of the two initial conditions of this equation, one can be fixed requiring that $f(R)/R \rightarrow 0 $ at early times (\ie  for large $R$) in order to recover GR; the second defines a one-parameter family of curves which all generate the given $H(a)$. Such parameter is conveniently chosen as (also see \cite{Starobinsky:2007hu}):
\be
\label{BHU}
 B_0 = \left(\frac{f_{RR}}{1 + f_R}R'\frac{H}{H'}\right)_0.
\ee
Here $f_R$ and $f_{RR}$ represent the first and second derivative of $f$ with respect to $R$, respectively. GR is represented by the special case $B_0 = 0$, so that $B_0$ effectively quantifies the deviation from GR at the present time. Furthermore, the gravitational stability condition is easily established as $B_0 > 0$. \\
The additional degrees of freedom of the $f(R)$ gravity introduce modifications in the Poisson equation and in the relation between the two gravitational potentials  \cite{Jain:2007yk}, which now read:
\be \label{poisson}
k^2 (\Phi - \Psi)  = 4\pi G_{\rm eff}(k,a) a^2 \rho_m \Delta_m 
\ee
and
\be
\Psi = (g(k,a)-1) \frac{k}{a H} (\Phi - \Psi).
\ee
The equation for the growth factor in MoG takes the form:
\be
\label{growthMG}
D'' + (2 + \frac{H'}{H})D'+  \frac{4\pi G_{\rm eff}(k,z) (g(k,z)-1)}{H^2}\rho_m D = 0.
\ee
Given $H(a)$, a MoG model is not completely defined unless $G_{\rm eff}(k,a)$, the effective Newton's Constant, and the metric ratio $ g(k,a)= (\Phi+\Psi)/(\Phi - \Psi)$, are known, contrary to the GR case. For the class of $f(R)$ theories under study, and in the sub-horizon, linear regime, Hu and Sawicki \cite{Hu:2007pj} have provided a fit to $G_{\rm eff}$ and $g$ as
\bea \label{g}
G_{\rm eff}(k,a)& = &G_{\rm eff}(a) = \frac{G_N}{1 + f_R}, \\ \nonumber
g(k,a) &=& \frac{g_{SH}(a) - \frac{1}{3} (0.71 \sqrt{B}(k/aH))^2}{1+ (0.71 \sqrt{B} (k/aH))^{2}},
\eea
where $B$ is the function appearing in Eq. (\ref{BHU}), and $g_{SH}(a)$ is the super-horizon metric ratio (see \cite{Song:2006ej,Hu:2007pj} for details). \\
The key result is that the $f(R)$ models all predict an enhancement in the growth rate of structure.
In fact, any positive value of $B_0$ gives rise to a negative $f_R$, so that the effective Newton constant is larger with respect to GR. Moreover, both terms in $g(k,a)$ have a negative sign, which induce further enhancement of matter clustering, as can be seen from Eq. (\ref{growthMG}).  
Since $G_{\rm eff}(k,z) = G_N/(1+ f_R)$ does not depend on $k$, the scale dependence of the growth factor can only arise from the second term of Eq. (\ref{g}). On large scales, the dominant term in the metric ratio is $g_{SH}(a)$, while for increasing values of $k$, the second term in the expression for $g(k,a)$ becomes important and tends to the constant value of $-1/3$ for large $k$.  The scale of the transition from scale-free to scale-dependent growth factor is $k/aH \simeq B^{-1/2}$; due to the asymptotic behavior of $g(k,a)$, the growth differs significantly from GR even for models with very small values of $B_0$, on sufficiently small scales. A mechanism to restore GR on scales of the galaxy and smaller is discussed in \cite{Hu:2007nk,Hu:2007pj}.  Fig. \ref{fig:growth}
shows $ d\ln(D)/d\ln(a)\Omega_m(a)^{\gamma} -1 $ for a few different values of $B_0$ and $k/a H_0$. For the GR case, $B_0 = 0$, $ \epsilon(k,a) - 1 \simeq 0$ with no scale dependence. 

\section{Measuring the growth of structure}

Peculiar velocities displace galaxies along the line of sight in redshift space and distort the power spectrum of galaxies observed  in redshift space. This effect  is known as linear redshift space distortion and was first derived by Kaiser \cite{Kaiser:1987qv}. In redshift-space, the power spectrum is amplified by a factor $(1+\beta \mu_{\mathbf{k}}^2)^2 $ over its real-space counterpart,
\be
P^s(\mathbf{k}) = (1+\beta \mu_{\mathbf{k}}^2)^2 P(k),
\ee
where $P^s(\mathbf{k})$ and $P(k)$ are redshift and real space power spectra, respectively, $\mu_{\mathbf{k}} = \hat{z}\cdot\hat{k}$ is the cosine of the angle between the wavevector $\hat{k}$ and the line of sight $\hat{z}$, and $\beta$ is the linear redshift-space distortion parameter defined as
\be
\beta(a) = \frac{1}{b} \frac{d \ln D}{d \ln a};
\ee
$b$ is the linear bias, which we assume to be independent of scale.  

Galaxy redshift surveys can be used to directly measure $\beta$, and, if the bias is known, the growth rate of perturbations. In fact,  the redshift space power spectrum can be decomposed into harmonics, whose relative amplitude depend on the growth rate of structure through $\beta$. The possibility of using such dependence in order to constrain dark energy properties has been explored in \cite{guzzo}; here we focus on the measurements of the scale dependence of $\beta$ as a smoking gun of Modified Gravity.\\
We assume that $\beta$ is obtained through the ratio of quadrupole to monopole moments of the redshift power spectrum \cite{Hamilton:1997zq}
\be
\frac{P_2(k)}{P_0(k)} = \frac{\frac{4}{3}\beta + \frac{4}{7}\beta^2}{1 + \frac{2}{3}\beta + \frac{1}{5}\beta^2},
\ee
and use the prescription in \cite{Feldman:1993ky} to get the errors in the above quantities:
\be
\frac{\sigma (P_i(k))}{P_i(k)} =  \left( \frac{(2\pi)^3 \int d^3{\bf r} \bar{n}^4({\bf r}) \psi^4({\bf r}) [1 + \frac{1}{\bar{n}({\bf r}) P_i(k)}]^2} { V_k \, \left[ \int d^3{\bf r} \bar{n}^2({\bf r}) \psi^2({\bf r}) \right]^2} \right)^{1/2}
\ee
where $\bar{n}({\bf r})$ is the mean galaxy density, $\psi$ is the weight function,  $V_k$ is the volume of the shell in $k$-space, and the index "i" assumes the values 0 and 2 for monopole and quadrupole, respectively. \\

The linear bias for a population of large-scale structure tracers can be estimated by  cross-correlating  the line-of-sight projected density of the tracer with a convergence map reconstructed by  CMB lensing techniques,  and comparing the resulting signal with theory. The weak lensing potential responsible for lensing the CMB can be written as the line-of-sight integral \cite{Bartelmann:1999yn}, 
\be
\phi(\nhat)=- \int  d\eta\frac{d_A(\eta_0-\eta)}{d_A(\eta_0) d_A(\eta) }[\Phi - \Psi](d_A(\eta) \nhat,\eta),
\ee
where $d_A(\eta)$ is the comoving angular diameter distance corresponding to the comoving distance $\eta$, and $\eta_0$ is the comoving distance to the last scattering surface. A quadratic combination of the measured CMB temperature and polarization \cite{Hu:2001kj,Okamoto:2003zw,Hirata:2002jy}  provides an estimator of the convergence field, $\kappa= \frac12 \nabla^2\phi$.
In this study, we have used the prescription of \cite{Hu:2001kj} to compute the expected noise power spectrum, $\nl^{\k\k}$, corresponding to the reconstructed convergence field by cross-correlating $\kappa$ with the projected fractional overdensity of the tracer,
\be
\Sigma(\nhat)=\int d\eta  W(\eta) b \,\delta_m(\eta \nhat,\eta)
\ee
where $W$ is the normalized tracer distribution function in comoving distance. We measure the cross-correlation spectrum:
\be
\cl^{\k-\Sigma}= \frac32 b\:  {\Omega_m}H_0^2 \int d\eta  \frac{ W(\eta)}{a(\eta)}P(\frac{\ell}{d_A},\eta)  \frac{d_A(\eta_0-\eta)}{d_A(\eta) d_A(\eta_0) }
\ee
where $P(k,\eta)$ is the matter power spectrum at the comoving distance $\eta$ and we have related the wavenumber $k$ to the multipole $\ell$ via the Limber approximation \cite{Limber}. The signal-to-noise ratio for such a cross-correlation can be estimated as \cite{Peiris:2000kb}, 
\be
\label{sncross}
\left(\frac{S}{N}\right)^2=\fsky\sum (2 \ell+1) \frac{\left(\cl^{\k-\Sigma}\right)^2}{\left(\cl^{\k\k}+N_l^{\k\k}\right)\left(\cl^{\Sigma\Sigma}+\nl^{\Sigma\Sigma}\right)}
\ee
where $\fsky$ is the fraction of sky over which the cross-correlation is performed. For tracer counts the noise is Poisson, and the power spectrum is given by,
$\nl^{\Sigma\Sigma}=1/{\hat{n}}$ where $\hat{n}$ is the number of tracer objects per steradian.
\par
\begin{figure}
\includegraphics[width=8.5cm]{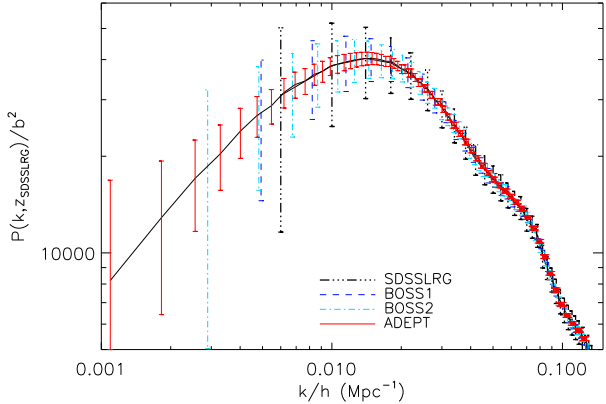}
\caption{Errors on $P(k)$, normalized to the SDSS-LRG median redshift (z=0.31) for all surveys.}
\label{fig:ps}
\end{figure}

Since the signal is proportional to the bias, $b$, the expected error on $b$ can be written as 
\be
\Delta b / b \simeq 1/(S/N) . 
\ee
 We consider three present and forthcoming redshift galaxy surveys: the SDSS LRG sample \cite{Tegmark:2006az}, its extension BOSS-LRG \cite{BOSS}, which we divide in two redshift bins, labeled as BOSS1 and BOSS2, and the proposed survey ADEPT \cite{ADEPT}.  Specifics of each experiment are listed in Table~\ref{bias_error}. In all cases we assume that $\beta$ and $b$ do not change significantly with redshift within a survey, so that the observed quantity is $\beta(k,z_{c})$, where $z_c$ is roughly the central redshift of the survey.
As a direct comparison the capabilities of the three galaxy surveys under examination, we show the real space matter power spectrum, normalized to the SDSS LRGs median redshift, with its errorbars in Fig. \ref{fig:ps}.
We also consider three possible CMB experiments:  a PLANCK-like CMB experiment with $65\%$ sky coverage and temperature and polarization sensitivities of 28 $\mu$K-arcmin and 57 $\mu$K-arcmin, respectively;  a next generation CMB survey based on using a camera similar to that on ACT or SPT with a polarimeter and a $\sim 3$ years observing program (labeled PACT) with  $65\%$ sky coverage and temperature and polarization sensitivities of 13 $\mu$K-arcmin and 18 $\mu$K-arcmin and  an ideal polarization experiment (labeled IDEAL), with  $65\%$ sky coverage and temperature and polarization sensitivities of 1 $\mu$K-arcmin and 1.4 $\mu$K-arcmin, respectively. The expected results from cross-correlation with the ADEPT and BOSS surveys, and the SDSS LRG are displayed in Table~\ref{bias_error}.
\begin{table}
  \center
  \begin{tabular}{@{\extracolsep{\fill}}lccccccc}
    \hline
    \hline
    Galaxy & $\hat{n}$ & $A/10^3$ & $z_c$ & b & CMB Expt. &  (S/N) & $\Delta b/b (\%)$\\
    Survey       &  &  & & & & & \\
    \hline
    \hline
    \hline
                  &        &       &        &       & PLANCK &  5.8  &17.3 \\
     SDSSLRG      & 12.4   & 3.8   &   0.31 &   2   & PACT & 11.4  & 8.8 \\
                  &        &       &        &       & IDEAL  & 20.4  & 4.9 \\
     \hline    
                  &        &       &        &       &    PLANCK & 10.8  & 9.3 \\
     BOSS1        &   40.  &  10   &   0.3  &   2   &    PACT & 25.5  & 3.9 \\
                  &        &       &        &       &    IDEAL  & 52.5  & 1.9 \\
     \hline
                  &        &       &        &       & PLANCK & 17.0  & 5.9 \\
     BOSS2        & 110.   & 10    &   0.6  &   2   & PACT & 39.4  & 2.5 \\
                  &        &       &        &       &IDEAL  & 78.2  & 1.3 \\
     \hline
                  &        &       &        &       & \small{PLANCK} & 52.8  & 1.9 \\
     ADEPT        &  3500  &  27   &   1.35  &   1   &         PACT & 107.5 & 0.9 \\
                  &        &       &        &       &         IDEAL  & 228.3 & 0.4 \\
     \hline
     \hline
  \end{tabular}
\caption{\label{bias_error} Predictions for the errors on bias from the cross-correlation studies described in the text. For each combination of experiments, we display the number of galaxies per square degree ($\bar{n}$); the area of overlap $(A)$, the signal-to-noise with which the cross correlation of tracer surface density with CMB-lensing can be extracted, (S/N), and the percentage error in the bias, $b$, for the tracer.}
\end{table}

\begin{figure*}
 \includegraphics[width=\linewidth]{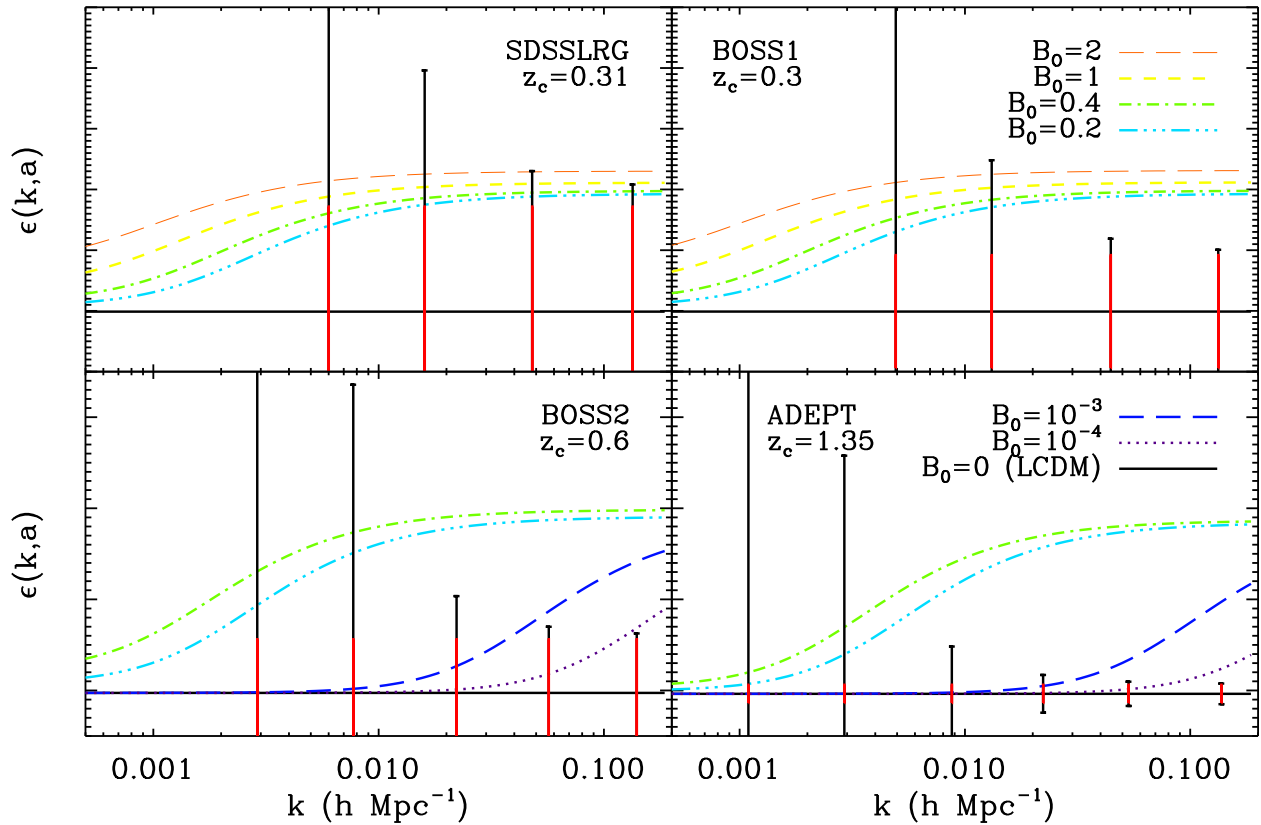} 
\caption{$\epsilon(k,z)$ for the four surveys, as a factor of $B_0$ and k. Total errorbars around the $\Lambda$CDM case are shown in black; the smaller red errorbars are from bias only.}
\label{fig:4panel}
\end{figure*}

\section{Results}
We show our main results in Fig. \ref{fig:4panel}. Errorbars are computed using Planck as the complementary CMB lensing survey for SDSS LRG and BOSS, and PACT for ADEPT. 
We summarize the current status of $\beta$ and bias measurements in Table~\ref{table1}, and add the expected errorbars on $\epsilon$ at two different scales, $k/h$ = 0.05 $\rm{Mpc}^{-1}$ and $k/h$ = 0.2 $\rm{Mpc}^{-1}$, from our analysis, for comparison.\\
The errors on bias are scale-independent and vary from 17$\%$ for SDSS LRG to $\sim$ 1 $\%$ for ADEPT. Errors on $\beta$ depend on scale; we bin our simulated data in bins in $k$ space of width $\Delta \ln k$ = $1/e$, and see that for all surveys errorbars decrease as we go to smaller scales.  The corresponding constraints on $\epsilon$ are as strong as a few percent for BOSS, and of the order of 1\% for ADEPT. 

The combined effect of smaller errors and of the asymptotic behavior of the growth factor, which induces a large deviation of $\epsilon(k,a)$ from its GR value on small scales, is that redshift galaxy surveys are more sensitive to the small-scale modification of gravity than to the large-scale one. Ultimately, the smallest observable value of $B_0$ will not be set by the capabilities of the survey, but by the breakdown of the linear regime assumption. Assuming $k \simeq 0.05$ Mpc$^{-1}$ as an upper limit, and for values of $aH$ corresponding to redshifts between 0 and 2, the smallest $B_0$ inducing scale-dependent growth is $B_0 \simeq 10^{-4}$. Such value is within reach of ADEPT; future experiments that detect redshifted 21 centimeter emission could probe even larger values of $k/aH$ in the linear regime.

\section{Conclusions and discussion}
We have built a null test parameter for General Relativity, $\epsilon(k,a)$, based on the consistency between expansion history and structure growth expected in GR. Such parameter can be expressed in terms of the combination $\Omega_{0m} h^2$, probed by the CMB experiments, the linear matter perturbations growth factor, probed by redshift galaxy surveys,  and the linear bias, probed by cross-correlation of the two. \\
We have predicted the achievable precision in the measurement of $\epsilon(k,a)$ for three redshift galaxy surveys, SDSS LRG, BOSS and ADEPT, together with Planck and a possible future CMB experiment, PACT. We have interpreted such result in the context of a one-parameter family of modified gravity theories, known as $f(R)$, which can give rise to cosmic acceleration. In such models, the matter clustering is enhanced on all scales with respect to the GR case, and the enhancement is largest on small scales. We concluded that the peculiar signatures of the $f(R)$ theories will be definitely detectable with a survey like ADEPT.\\
More generally, any detection of deviation of $\epsilon$ from zero that was not due to some observational systematic would be a signature of truly novel physics with enhanced growth, pointing either to non-GR physics or to unexpected properties of dark energy: dark energy models with a non-zero sound speed are characterized by an oscillatory behavior of the growth \cite{DeDeo:2003te}, and scalar field dark energy suppresses growth on large scales \cite{Unnikrishnan:2008qe}. Similarly, massive neutrinos suppress $\epsilon(k,a)$ on scales below the neutrino free streaming scale (see \cite{Lesgourgues:2006nd} for review). \\

We warmly thank E. Aubourg, C. Hirata, W. Hu, R. H. Lupton, M. A. Strauss and L. Verde for useful suggestions. DNS thanks the APC in Paris for its hospitality. This work was supported by NSF grant AST-0707731, the NSF PIRE program and the NASA LTSA program. 


\
\vspace{0pt}
\begin{table*}[!t]
\begin{center}
\caption{Currently available data for measurements of $\epsilon$ through $\beta$ and $b$ (from \cite{Nesseris:2007pa}, with the addition of the measurement reported in \cite{guzzo}), and comparison with our predictions. Only the error coming from uncertainties in $\beta$ and $b$ is considered.}
\label{table1}
\resizebox{\textwidth}{!}{%
\begin{tabular}{cccccccccc}
\hline \hline\\
\hspace{6pt}   $z$  &\hspace{6pt} $\beta$ & \hspace{6pt} $b$ &\hspace{3pt}  $ \Delta \epsilon/ \epsilon (\%) $ &\hspace{6pt}  \text{Ref.} & & $z$ &  $\Delta \epsilon/ \epsilon  (\%)$ & $\Delta \epsilon/ \epsilon (\%)$ & COMBINATION \\
& & & & & & & $  k \, = \,0.05 \, h/\rm{Mpc}$  & $k \, = \,0.2 \, h/\rm{Mpc}  $ & OF SURVEYS \\
\hline\\
\hspace{6pt}   0.15 &\hspace{6pt} $0.49 \pm 0.09$         &\hspace{6pt} $1.04\pm 0.11$ &\hspace{6pt}  $21.5$ & \cite{Hawkins:2002sg},\cite{Verde:2001sf} & \hspace{6pt} & 0.3 & 22.0& 10.1& BOSS1 + Planck \\ 
\hspace{6pt}   0.35 &\hspace{6pt} $0.31 \pm 0.04$         &\hspace{6pt} $2.25\pm 0.08$ &\hspace{6pt}  $25.7$ & \cite{Tegmark:2006az} & \hspace{6pt} & 0.31 & 39.5& 21.0& SDSS LRG + Planck \\
\hspace{6pt}   0.55 &\hspace{6pt} $0.45 \pm 0.05$         &\hspace{6pt} $1.66\pm 0.35$ &\hspace{6pt}  $24.0$ & \cite{Ross:2006me} & \hspace{6pt} & 0.5 & 9.3& 5.5& BOSS + Planck\\
\hspace{6pt}   0.77  &\hspace{6pt}$0.70 \pm 0.26$         &\hspace{6pt}$1.3 \pm 0.1$             &\hspace{6pt}  $39.6$ & \cite{guzzo} &\hspace{6pt}  & 0.6 & 10.6& 6.5& BOSS2 + Planck \\
\hspace{6pt}   1.4  &\hspace{6pt} $0.60 ^{+0.14}_{-0.11}$ &\hspace{6pt} $1.5 \pm 0.20$ &\hspace{6pt}  $27.7$ & \cite{daAngela:2006mf}  &\hspace{6pt} & 1.35 & 2.1 & 1.1& ADEPT + PACT\\
\hspace{6pt}   3.0  &\hspace{6pt} $-$                     &\hspace{6pt}$-$             &\hspace{6pt}  $19.9$ & \cite{McDonald:2004xn} & \hspace{6pt}& & & & \\
\hline \hline \\
\end{tabular}}
\end{center}
\end{table*}

\end{document}